\documentclass[10pt,twoside,BCOR7mm,DIV15,headinclude,footexclude,
               cleardoubleempty,idxtotoc]
{scrartcl}

\usepackage{natbib}
\usepackage[font=small,labelfont=bf]{caption}
\usepackage[english]{babel}
\usepackage{graphicx}
\usepackage{hyperref}
\usepackage{scrpage2}
\usepackage{ifthen}
\usepackage{booktabs}
\usepackage{amsmath}
\usepackage{amssymb}
\usepackage{multicol}
\usepackage{float}
\usepackage{hyperref}

\hypersetup{breaklinks=true
,colorlinks=true,linkcolor=black,urlcolor=blue
,citecolor=black}

\addto\captionsenglish{%
}

\makeatletter
\renewcommand{\@biblabel}[1]{}
\renewcommand{\@cite}[2]{%
{#1\ifthenelse{\boolean{@tempswa}}{,#2}{}}}
\makeatother
\ofoot{\thepage}
\ifoot{}

\setcapindent{0em}
\setheadsepline{1pt}

\setkomafont{pagehead}{\normalfont\sffamily}
\setkomafont{pagenumber}{\normalfont\rmfamily}

\makeatletter
\newcommand{\listofcontributions}{\@starttoc{con}}

\newcommand{\l@contribution} {\@dottedtocline{1}{1.5em}{2.3em}}
\makeatother

\newenvironment{contribution}{
\setcounter{section}{0}
\setcounter{figure}{0}
\setcounter{table}{0}
}{
\newpage
\lehead{}
\rohead{}
}

\def\rso{$\,{\rm R}_{\odot}\,$}                
\def\mso{$\,{\rm M}_{\odot}\,$}

\begin{document}

\setlength{\baselineskip}{2.5ex}

\begin{contribution}

\lehead{D.\ Sanyal, T.\ J.\ Moriya \& N.\ Langer}

\rohead{Envelope inflation in WR stars}

\begin{center}
{\LARGE \bf Envelope inflation in Wolf-Rayet stars and
extended supernova shock breakout signals}\\
\medskip

{\it\bf D.\ Sanyal, T.\ J.\ Moriya \& N.\ Langer}\\

{\it Argelander-Institut f\"ur Astronomie, Universit\"at Bonn, Germany}\\

\begin{abstract}
Massive, luminous stars reaching the Eddington limit in their interiors develop 
very dilute, extended envelopes. This effect is called envelope {\it inflation}. 
If the progenitors of Type Ib/c supernovae, which are believed to be  
Wolf-Rayet (WR) stars, have inflated envelopes then the shock breakout 
signals diffuse in them and can extend their rise times significantly. We show 
that our inflated, hydrogen-free, WR stellar models with a radius of $\sim$\rso can 
have shock breakout signals longer than $\sim 60\,s$. The puzzlingly long shock breakout 
signal observed in the Type Ib SN 2008D can be explained by an inflated progenitor envelope, and 
more such events might argue in favour of existence of inflated envelopes in general. 
\end{abstract}
\end{center}

\begin{multicols}{2}

\section{Stellar envelope inflation}
Inflation refers to the extremely dilute, loosely-bound envelopes that 
massive, luminous stars develop in the course of their evolution 
\citep[][Langer et al., this volume]{ishii99,petrovic_2006,sanyal2015}.
A typical inflated envelope is shown in Fig.~\ref{deb:fig:inflation_eg}, which 
corresponds to an evolved hydrogen-free 7.8 \mso model burning carbon in the core. 
Such ubiquitous hydrostatic structures result from stellar layers reaching the 
Eddington limit in the interior of the star, facilitated by the peak in  
opacity at $T\approx 200\,000$ K, the so-called iron opacity bump. 
For more details on the appropriate definition of the Eddington limit, we 
refer to \citet[][]{sanyal2015} and Langer et al. (this volume).

\begin{figure}[H]
\begin{center}
\includegraphics[width=0.7\columnwidth,angle=270]{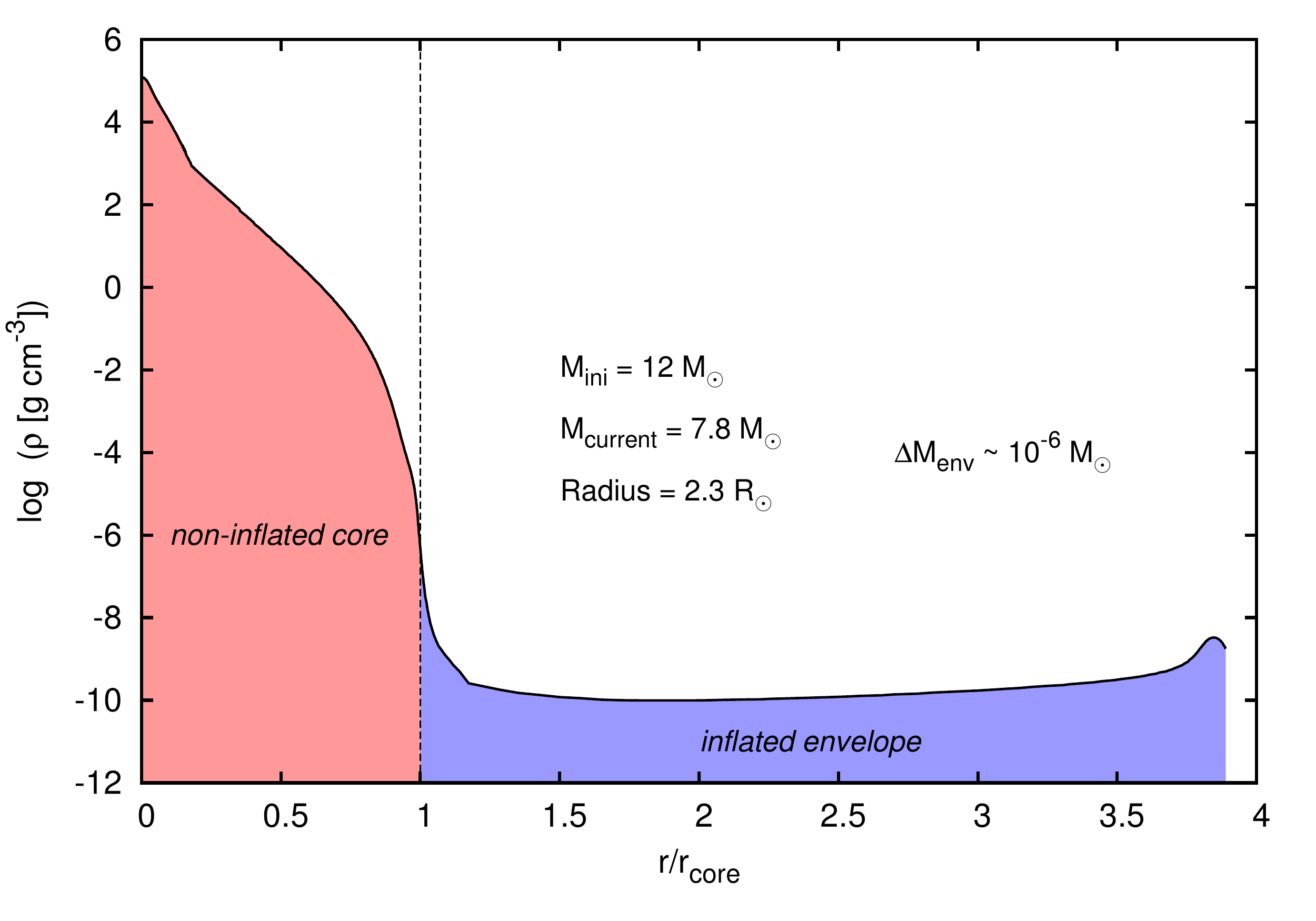}
\caption{Density profile of a 7.8 \mso hydrogen-free stellar model 
with $T_{\rm eff} = 82\,700$ K and $\log (L/L_{\odot} ) = 5.35$, showing an
inflated envelope (shaded blue) and a density inversion. The radial co-ordinate 
has been normalised by the non-inflated core radius ($r_{\rm core}$) of $0.6$ \rso.
\label{deb:fig:inflation_eg}}
\end{center}
\end{figure}

Stellar layers approaching the Eddington luminosity become 
convectively unstable \citep{langer97} but convection, of course, 
does not contribute to the radiative acceleration. However, 
the fraction of the total flux transported by convection 
in these WR stars, according to standard MLT, 
is negligibly small. Hence, whether or not 
the layers will hit the Eddington limit and inflate the envelope,  
does not depends critically on the theory of convective energy transport 
used to model the stars. 

As we can see from Fig.~\ref{deb:fig:inflation_eg}, 
the mass contained in the inflated envelope is a tiny 
fraction of the stellar mass, $\Delta M_{\rm env}\approx 10^{-6}\,$\mso. 
However, for cool supergiants, the mass in the 
inflated envelopes can be as high as a few solar masses, see 
\citet{sanyal2015} for details.

\section{Stellar models of WR stars}

We modelled WR stars as non-rotating, hydrogen-free helium stars and 
evolved them through core-helium buring to the onset of core-carbon 
burning, using the 1-D stellar evolution code 
BEC \citep[see][for details]{heger2000,ines2011,moriya2015}.
We computed two sets of models, with initial masses 10 \mso and 12 \mso, 
and a metal composition that of the Sun \citep{ines2011}. In particular, we used the OPAL 
opacity tables \citep{ir96} for our computations and used the 
standard Mixing-Length-Theory (MLT) to model convective zones, 
with $\alpha_{\rm MLT}=1.5$. The mass-loss rate prescription from 
\citet{nugis&lamers2000} was applied. The resulting structure 
(e.g. Fig.~\ref{deb:fig:inflation_eg}) will resemble a SN Type Ib progenitor 
because no significant changes in the stellar structure is expected 
in the remaining lifetime until the explosion. 

The evolutionary tracks of the aforementioned models, 
in the H-R diagram, is shown in Fig.~\ref{deb:fig:hrd}.
Both the helium zero age main sequence models 
are marginally inflated. As a consequence of the applied
mass loss, they decrease in luminosity during
the core helium-burning phase. After helium is exhausted
in the core, the stars begin to contract and become hotter and
brighter. The models are hardly inflated during this phase  because
the Fe-group opacity bump is only partially contained inside
the stars. These are probably related to the WO stars (cf. Tramper et al., this volume). 
When the helium-shell ignites, the evolutionary
tracks eventually move toward cooler temperatures because of
the mirror principle. As the models
become cooler, they become significantly inflated and exhibit
a pronounced core-halo structure, as shown in 
Fig.~\ref{deb:fig:dens} \citep[cf.\,][]{moriya2015}.


\section{Shock breakout in inflated stellar envelopes}

The shock breakout occurs when the dynamical timescale 
of the shock propagation in the unshocked envelope, i.e. $t_{\rm dyn}\simeq \Delta R/v_{\rm sh}$ 
becomes comparable to the diffusion timescale 
in the envelope, i.e. $t_{\rm diff} \simeq \tau \Delta R/c$, where $v_{\rm sh}$ is the shock 
velocity and $\tau$ is the optical depth in the remaining unshocked envelope \citep{weaver1976}. 
Therefore, the shock breakout condition can be expressed as: $\tau\simeq c/v_{\rm sh}$.

\begin{figure}[H]
\begin{center}
\includegraphics[width=0.7\columnwidth,angle=270]{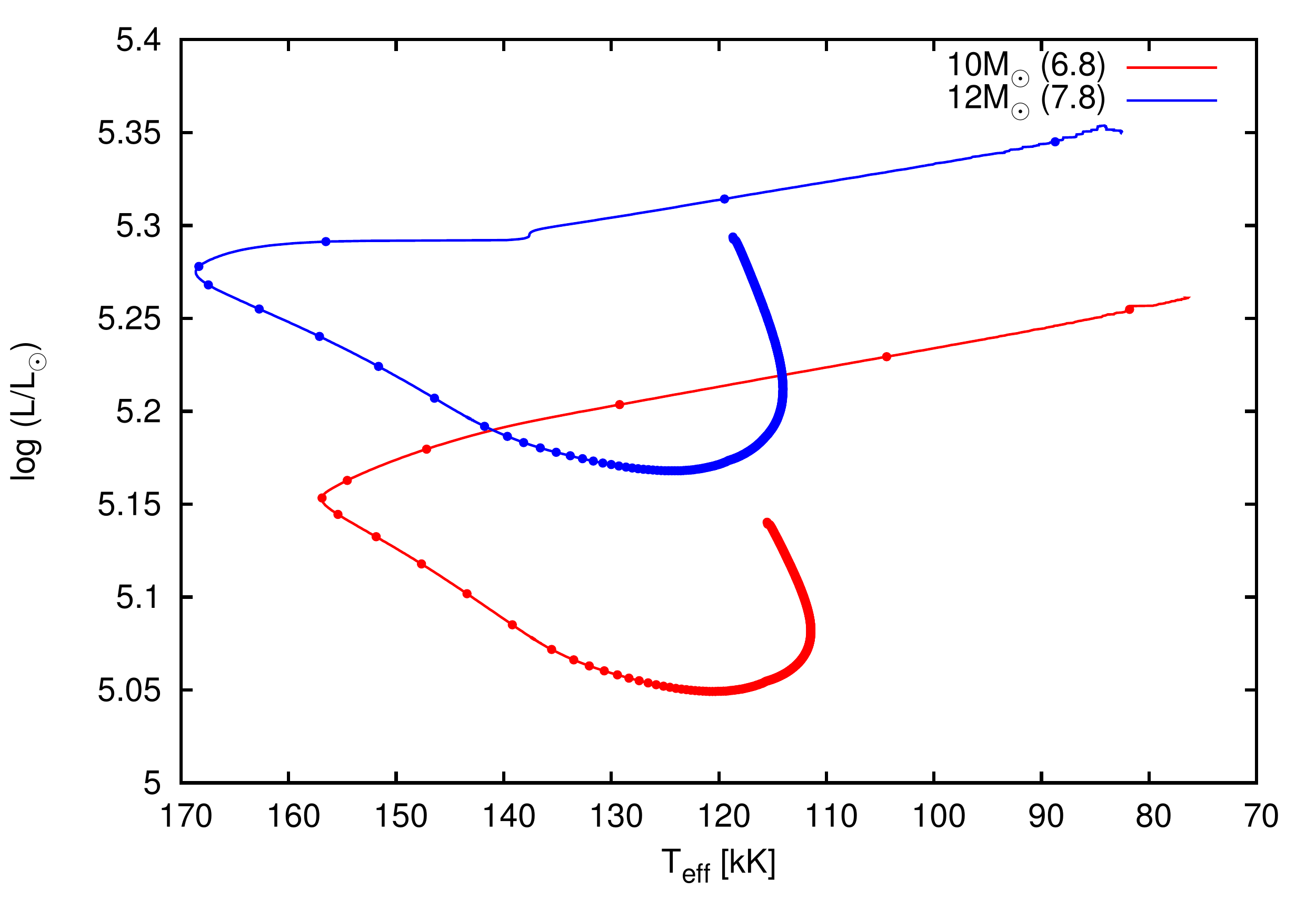}
\caption{Stellar evolutionary tracks of our $10$\mso and $12$\mso solar-metallicity helium-star models 
from the He-ZAMS until core-carbon ignition.
\label{deb:fig:hrd}}
\end{center}
\end{figure}

For a model in which the shock breakout condition is fulfilled 
very close to the stellar surface, $\Delta R$ is very small, such that 
$t_{\rm diff}\ll t_{\rm lc}$, where $t_{\rm lc}\simeq R_{\star}/c$ is the light-crossing time. 
Assuming a shock velocity of $20\,000\, {\rm km\,s^{-1}}$, in the case of the polytropic model 
shown in Fig.~ \ref{deb:fig:dens}, the light-crossing time $t_{\rm lc}$ is 5.6\,s whereas the diffusion 
time is only 0.5 s, meaning that the shock breakout duration is determined by $t_{\rm lc}$.

On the other hand, in an inflated stellar model, the shock breakout happens relatively deep 
inside the envelope, making $\Delta R$ quite large, such that $t_{\rm diff}\gg t_{\rm lc}$. 
In Fig.~\ref{deb:fig:tau}, we plot the optical depth in the interior of our stellar 
models and show that $\Delta R$ can be as large as $\sim $\rso.
In such a situation, the rise time of the shock breakout signals
 will be determined by the diffusion time. The 
subsequent light curve is expected to decline exponentially with
an e-folding timescale of the diffusion time due to the photon
diffusion in the shocked envelope \citep{moriya2015}. 
Table \ref{Debashis:table:timescales} summarises the different timescales 
involved in inflated and un-inflated models.

\begin{figure}[H]
\begin{center}
\includegraphics[width=0.7\columnwidth,angle=270]{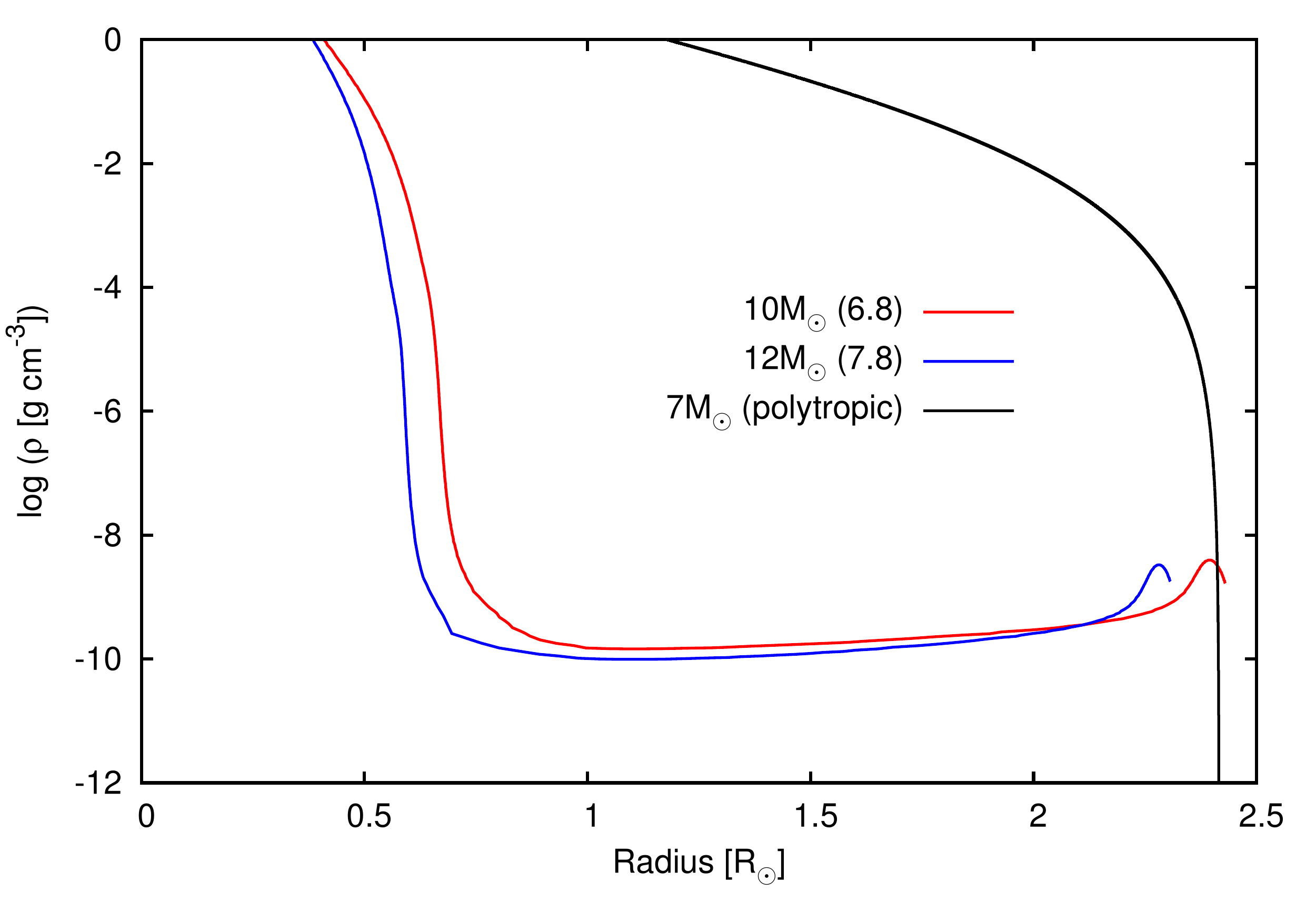}
\caption{Density structure of our last computed models, with initial masses of 
$10$ \mso and $12$ \mso. The final masses are indicated in parentheses, 
 in units of a solar mass. The density structure of a 7 \mso polytropic model 
(polytropic index = 3) without an inflated envelope is also shown.
\label{deb:fig:dens}}
\end{center}
\end{figure}

\textit{SN 2008D:} The X-ray light curve of the Type Ib SN 2008D was serendipitously observed 
by \citet{soderberg2008} with the \textit{SWIFT} satellite, and is the only direct detection of 
a SN shock breakout so far. The total 
shock breakout duration was $300$ seconds with a rise time of $\sim 60$ seconds. 
If we assume the shock brekout duration was dominated by $t_{\rm lc}$, as is done in 
the literature, then it implies a WR progenitor radius of $130$ \rso, which is much 
larger than predicted by stellar evolutionary models. On the other hand, our inflated 
model with a radius of $\sim 2$ \rso can naturally explain the long rise time of SN 2008D. 
The estimated SN ejecta mass is $3-7$ \mso \citep{soderberg2008,mazzali2008,bersten2013} which is consistent with the 
final masses of our models (after subtracting the remnant mass of the neutron star). 
The mass-loss rates of our models ($7\times 10^{-6}\,{\rm M_{\odot}\,yr^{-1} }$)
are also consistent with that estimated from radio observations
\citep[$7\times 10^{-6}\,{\rm M_{\odot}\,yr^{-1} }$,][]{soderberg2008}.

The shock velocity at shock breakout has been suggested to be $\sim 30\,000\, {\rm km\,s^{-1}}$ 
for WR stars \citep[e.g.,][]{nakar2010}, but these studies assume a steeply declining density 
profile as in the case of the polytropic model. Hydrodynamic modelling of shock propagation 
in inflated envelopes is needed to estimate the shock velocity at breakout.

There have been other models proposed for explaining the long shock breakout signal of 
SN 2008D, like the thick wind model by \citet{svirski2014} which assumes a 
high mass-loss rate of $\sim 10^{-4}\,{\rm M_{\odot}\,yr^{-1}}$ a few days prior to the 
explosion, and the supernova ejecta expanding through the optically-thick wind.

\begin{table}[H] 
\begin{center} 
\captionabove{A comparison of the properties and the relevant timescales of our computed models. The 
shock breakout rise time of SN 2008D was $\sim 60$ seconds.} 
\label{Debashis:table:timescales}
\begin{tabular}{lccc}
\toprule
Mass (\mso)   & $t_{\rm lc}$ (s) & $t_{\rm diff}$ (s) & $R_{\star}$ (\rso)\\
\midrule 
6.8   & 5.67 & 14 & 2.43\\
7.8 & 5.38 & 50 & 2.31\\
7.0 (Polytropic)  & 5.60 & 0.5 & 2.40\\
\bottomrule
\end{tabular}
\end{center}
\end{table}

\begin{figure}[H]
\begin{center}
\includegraphics[width=0.7\columnwidth,angle=270]{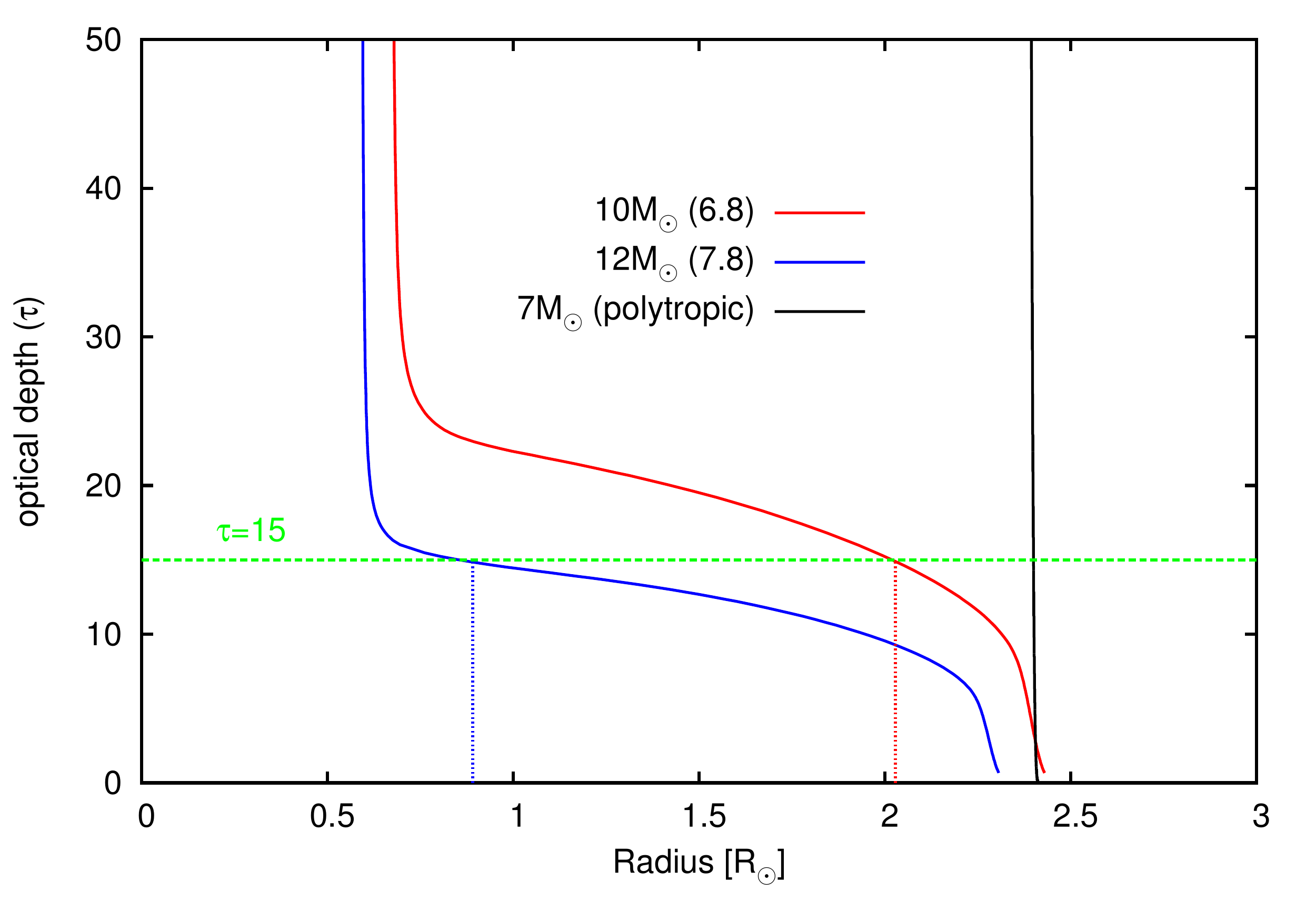}
\caption{The run of optical depth $\tau$ in our models. For computing $\tau$ in the 
polytropic model, a constant opacity coefficient of $0.2\,{\rm cm^2\,g^{-1}}$ was used. The 
green dashed line at $\tau=15$ is drawn for convenience. The blue and red dotted lines show where in the 
stellar envelope shock breakout happens.
\label{deb:fig:tau}}
\end{center}
\end{figure}
\section{Conclusions}

We have shown that if WR type supernova progenitors have 
inflated envelopes, then the rise times of supernova shock breakout
signals can be extended because the shock breakout can then 
occur within the low-density envelopes.
The long diffusion time of the inflated envelopes makes
the shock breakout rise times long. 
Even if a SN progenitor has a radius on the order
of the solar radius whose light-crossing time is a few seconds,
the rise time of the shock breakout signals can be $\sim 60$ s 
because of the inflated envelope. Our inflated model can simultaneously explain 
the mysterious long shock breakout rise time, the ejecta mass, the 
mass-loss rate estimate right before explosion, and the shock velocity  
in SN 2008D.

The inflated envelope is a generic feature of luminous, mass stars 
and can have an array of observational consequences. It leads to cooler 
effective temperatures, higher spin-down rates in rotating stars, and 
less massive supernova progenitors. It has been argued in the literature 
\citep{graefener_2012} that inflated 
WR envelopes can address the WR radius problem. 
A recent study of galactic OB stars by \citet{castro_2014} 
indicates that inflated envelopes do exist in nature. Furthermore, core-hydrogen burning 
stars evolving to 
temperatures below $\sim 8000$ K can be expected to have massive inflated envelopes and might 
be related to LBV giant eruptions \citep{sanyal2015}.
%

\bibliographystyle{aa} 
\bibliography{myarticle}

\end{multicols}

\end{contribution}


\end{document}